\documentstyle[12pt]{article}

\newcommand{\be}{\begin{equation}}
\newcommand{\ee}{\end{equation}}
\newcommand{\ba}{\begin{eqnarray}}
\newcommand{\ea}{\end{eqnarray}}
\newcommand{\ban}{\begin{eqnarray*}}
\newcommand{\ean}{\end{eqnarray*}}
\newcommand{\n}{\nonumber \\}

\newcommand{\eq}[1]{(\ref{#1})}
\newcommand{\sfrac}[2]{{\textstyle \frac{#1}{#2}}}
\newcommand{\ignore}[1]{}
\newcommand{\Romannumeral}[1]{\uppercase\expandafter{\romannumeral#1}}

\begin{document}

%%%%%%%%%%%%%%%%%%%%%%%%%%%%%%%%%%%%%%%%%%%%%%%%%%%%%%%%%%%
%                                                         %
%  Title page                                             %
%                                                         %
%%%%%%%%%%%%%%%%%%%%%%%%%%%%%%%%%%%%%%%%%%%%%%%%%%%%%%%%%%%
\renewcommand{\thefootnote}{\fnsymbol{footnote}}
\font\csc=cmcsc10 scaled\magstep1
{\baselineskip=14pt
 \rightline{
 \vbox{\hbox{YITP-96-1}
       \hbox{DPSU-96-1}
       \hbox{January 1996}
}}}

\vskip 10mm
\begin{center}
{\large\bf Reflection K-Matrices of the 19-Vertex Model\\
and\\ 
\vskip 1.6mm
$XXZ$ Spin-1 Chain with General Boundary Terms}

\vspace{10mm}

{\csc Takeo INAMI}$^a$,\setcounter{footnote} 
{0}\renewcommand{\thefootnote}{\arabic{footnote}}
{\csc Satoru ODAKE}$^{b,}$\footnote{
     e-mail address : odake@yukawa.kyoto-u.ac.jp.}
and 
{\csc Yao-Zhong ZHANG}$^{a,}$\footnote{
      Address after June 6, 1996 :
      Department of Mathematics, University of Queenland, Brisbane,
      Qld 4072, Australia;
      e-mail address : yzzhang@yukawa.kyoto-u.ac.jp,
      yzz@maths.uq.oz.au.}

\vskip3mm
{\baselineskip=15pt
\it\vskip.25in
  $^a$Yukawa Institute for Theoretical Physics \\
  Kyoto University, Kyoto 606, Japan \\
\vskip.1in
  $^b$Department of Physics, Faculty of Science \\
  Shinshu University, Matsumoto 390, Japan
}
\end{center}

\vspace{4mm}

\begin{abstract}
{%\baselineskip 15pt
We derive and classify all solutions of the boundary Yang-Baxter equation
(or the reflection equation) for the 19-vertex model associated with 
$U_q(\widehat{sl_2})$. Integrable $XXZ$ spin-1 chain hamiltonian with general
boundary interactions is also obtained.
}
\end{abstract}

\vskip.4in
hep-th/9601049
\renewcommand{\thefootnote}{\arabic{footnote}}

\newcommand{\sect}[1]{\setcounter{equation}{0}\section{#1}}
\renewcommand{\theequation}{\thesection.\arabic{equation}}

\newpage
%\pagestyle{plain}
%\pagenumbering{arabic}
%%%%%%%%%%%%%%%%%%%%%%%%%%%%%%%%%%%%%%%%%%%%%%%%%%%%%%%%%%%
%                                                         %
%  1. Introduction                                        %
%                                                         %
%%%%%%%%%%%%%%%%%%%%%%%%%%%%%%%%%%%%%%%%%%%%%%%%%%%%%%%%%%%
\sect{Introduction}

Many problems in particle physics and condensed matter physics can be
mapped to analogous problems in integrable quantum
field theories and/or solvable lattice models in two dimensions by 
employing bold simplification or approximation. Understanding of integrable
models has helped us gain insight into non-perturbative properties
of four-dimensional quantum field theories, such as chiral symmetry
breaking and confinement.  

In some physical systems,  effects of boundaries are very important
and hence for physical application we have to extend
integrable models to cases with boundaries. Methods for
constructing integrable models with boundaries have recently been
developed \cite{Che84,Skl88,Gho94,Car89}. One systematic approach lies in
finding the reflection matrix (or K-matrix) obeying the boundary 
Yang-Baxter equation (BYBE) (or reflection equation). Only few models with
integrable boundary terms in full generality have so far been constructed:
(super) sine-Gordon theory \cite{Gho94,Ina95a}, massive Thirring model
\cite{Ina95b}, affine Toda field theories \cite{Cor94},
$XXZ$ (and $XXX$) \cite{deV93,Jim95} and $XYZ$ spin-$\frac12$ 
chains \cite{Hou93}.  

Quantum spin chains have played important roles in many aspects of physics
such as strongly correlated electron systems. Recently
there has been experimental interest in quasi one-dimensional systems of
molecules which mimic spin chains of higher spin $s$. To our
knowledge, however, for spin
chain with $s\geq 1$ only boundary terms corresponding to diagonal
K-matrices are known (see, e.g., \cite{Mez90}).

In this paper we consider the 19-vertex model which is 
associated with the spin-1 representation
of $U_q(\widehat{sl_2})$.  This model is
the simplest extension of the six-vertex model. We construct and classify
all reflection K-matrices of this model. From these
K-matrices we derive the general integrable boundary terms for the
$XXZ$ spin-1 chain. As a special case we reproduce the hamiltonian
obtained in \cite{Mez90} by using the diagonal K-matrix.

%%%%%%%%%%%%%%%%%%%%%%%%%%%%%%%%%%%%%%%%%%%%%%%%%%%%%%%%%%%
%                                                         %
%  2. Nineteen-vertex Model                               %
%                                                         %
%%%%%%%%%%%%%%%%%%%%%%%%%%%%%%%%%%%%%%%%%%%%%%%%%%%%%%%%%%%
\sect{Nineteen-vertex Model}

The hamiltonian of the $XXZ$ spin-1 chain is derived from the transfer matrix
of the 19-vertex model. This vertex model is defined in terms of the
Boltzmann weights given by the R-matrix
of spin-1 representation of $U_q(\widehat{sl_2})$. It is the trigonometric
solution of the Yang-Baxter equation (YBE):
\be
  \Bigl(1\otimes\check{R}(u-u')\Bigr)
  \Bigl(\check{R}(u)\otimes 1\Bigr)
  \Bigl(1\otimes\check{R}(u')\Bigr)
  =
  \Bigl(\check{R}(u')\otimes 1\Bigr)
  \Bigl(1\otimes\check{R}(u)\Bigr)
  \Bigl(\check{R}(u-u')\otimes 1\Bigr),
  \label{YBE}
\ee
where $\check{R}(u)=PR(u)$ and $P$ is the transposition matrix:
$P\;v_1\otimes v_2=v_2\otimes v_1$. $R(u)$  is a $9\times 9$ matrix with 
19 non-zero entries \cite{rZF}:
\be
  R(u)=\rho(u)
  \left(
  \begin{array}{ccccccccc}
    a_1&&&&&&&&      \\
    &a_2&&a_3&&&&&   \\
    &&a_4&&a_5&&a_6&&\\
    &a_3&&a_2&&&&&   \\
    &&a_5&&a_7&&a_5&&\\
    &&&&&a_2&&a_3&   \\
    &&a_6&&a_5&&a_4&&\\
    &&&&&a_3&&a_2&   \\
    &&&&&&&&a_1
  \end{array}
  \right) \qquad
  \begin{array}{rcl}
    a_1&\!\!=\!\!&\sin(u+2\eta)\\
    a_2&\!\!=\!\!&\sin u\\
    a_3&\!\!=\!\!&\sin 2\eta\\
    a_4&\!\!=\!\!&\displaystyle\frac{\sin u\sin(u-\eta)}{\sin(u+\eta)}\\
    a_5&\!\!=\!\!&\displaystyle\frac{\sin 2\eta\sin u}{\sin(u+\eta)}\\
    a_6&\!\!=\!\!&\displaystyle\frac{\sin\eta \sin 2\eta}{\sin(u+\eta)}\\
    a_7&\!\!=\!\!&a_6+\sin u,
  \end{array}
  \label{R}
\ee
where $\rho(u)\not\equiv 0$ is an arbitrary function.
If we choose $\rho(u)=\rho_\pm(u)$, 
\be
  \rho_{\pm}(u)=\frac{1}{\sin(2\eta\pm u)},
  \label{rho}
\ee
then $R$ satisfies
\be
  \check{R}(u)\check{R}(-u)=1,\quad \check{R}(0)=1.
\ee
In the case of periodic boundary condition, the YBE implies a commuting
family of transfer matrices and hence the integrability of the model.

%%%%%%%%%%%%%%%%%%%%%%%%%%%%%%%%%%%%%%%%%%%%%%%%%%%%%%%%%%%
%                                                         %
%  3. K-matrix                                          %
%                                                         %
%%%%%%%%%%%%%%%%%%%%%%%%%%%%%%%%%%%%%%%%%%%%%%%%%%%%%%%%%%%
\sect{Reflection K-matrices}

We now consider the 19-vertex model with boundaries. As shown by Sklyanin
\cite{Skl88}, integrable models with boundaries can be constructed out of
a pair of reflection 
K-matrices $K_\pm(u)$ in addition to $R(u)$. $K_+(u)$ and $K_-(u)$
describe the effects of the presence of boundaries at the left and the right, 
respectively, and they both obey the BYBE.

Let us construct the K-matrix which satisfies the BYBE:
\be
  R(u-u')\Bigl(K(u)\otimes 1\Bigr)
  R(u+u')\Bigl(1\otimes K(u')\Bigr)
  =
  \Bigl(1\otimes K(u')\Bigr)R(u+u')
  \Bigl(K(u)\otimes 1\Bigr)R(u-u').
  \label{BYBE}
\ee
%The BYBE implies a commuting family of transfer matrices $t(u)$, which are 
%now defined using the K-matrices and the monodromy matrices $T(u)$ as
%\be
%t(u)={\rm tr}_0\left[K_+(u)T(u)K_-(u)T^{-1}(-u)\right]
%\ee
%where $T(u)=R_{N0}(u)\cdots R_{10}(u)$.  The commuting properties of $t(u)$,
%$[t(u), t(u')]=0$, follows from the YBE and the BYBE.
In the present case, $R(u)$ is given in \eq{R} and
$K(u)$ can be parametrized as
\be
  K(u)=\rho^K(u)
  \left(
  \begin{array}{ccc}
    x_1(u)&y_1(u)&z(u)\\
    \tilde{y}_1(u)&x_2(u)&y_2(u)\\
    \tilde{z}(u)&\tilde{y}_2(u)&x_3(u)
  \end{array}
  \right),
  \label{K}
\ee
where $\rho^K(u)\not\equiv 0$ is an arbitrary function.

The BYBE consists of 81 equations, many of which are dependent.
To solve them, we proceed in a pragmatic fashion. 
The whole procedure is quite involved and requires a bit of
computer work. Here we only present the final result. More details of
the derivation are relegated to the Appendix.

We pick up a few 
easy-looking and independent equations. They are functional equations and
can be reduced to first order
differential equations or algebraic equations. The solutions to these equations
contain several arbitrary parameters. 
We then substitute these partial solutions
back into the BYBE, which gives a set of equations with the number
of equations now
less than 81 (since some equations are already satisfied). From this set
of equations we pick up some 
simple-looking ones, solve them and determine the
relations among the parameters. Repeating this process for a couple of
times, we end up with the solutions with three arbitrary parameters
satisfying all 81 functional equations. The solutions obtained in this way
exhaust all solutions of \eq{BYBE} (for the 19-vertex model).

All solutions can be classified into the following three cases:
\vskip.1in
\noindent {\bf Class (A): $\sin 2\eta\neq 0$.}
\ba
  y_1(u)&\!\!\!=\!\!\!&\mu\sin(\zeta-\sfrac{\eta}{2}+u)\sin 2u,\quad
  \tilde{y}_1(u)=\tilde{\mu}\sin(\zeta-\sfrac{\eta}{2}+u)\sin 2u,\n
  y_2(u)&\!\!\!=\!\!\!&\mu\sin(\zeta+\sfrac{\eta}{2}-u)\sin 2u,\quad
  \tilde{y}_2(u)=\tilde{\mu}\sin(\zeta+\sfrac{\eta}{2}-u)\sin 2u,\n
  z(u)&\!\!\!=\!\!\!&\mu^2\frac{\sin(\eta-2u)}{2\cos\eta}\sin 2u,
  \qquad\;
  \tilde{z}(u)=\tilde{\mu}^2\frac{\sin(\eta-2u)}{2\cos\eta}\sin 2u,\n
  x_1(u)&\!\!\!=\!\!\!&
  \sin(\sfrac{\eta}{2}+\zeta+u)\sin(\sfrac{\eta}{2}-\zeta-u)
  +\mu\tilde{\mu}\frac{\sin(\eta-2u)}{2\cos\eta}\sin\eta, 
  \label{Kanswer}\\
  x_2(u)&\!\!\!=\!\!\!&
  \sin(\sfrac{\eta}{2}+\zeta-u)\sin(\sfrac{\eta}{2}-\zeta-u)
  +\mu\tilde{\mu}\frac{\sin(\eta-2u)}{2\cos\eta}\sin(\eta+2u), \n
  x_3(u)&\!\!\!=\!\!\!&
  \sin(\sfrac{\eta}{2}+\zeta-u)\sin(\sfrac{\eta}{2}-\zeta+u)
  +\mu\tilde{\mu}\frac{\sin(\eta-2u)}{2\cos\eta}\sin\eta, \nonumber
\ea
where $\zeta,\mu,\tilde{\mu},$ are arbitrary constants.
If we choose $\rho^K(u)=\rho^K_{\pm}(u)$ with
\be
  \rho^K_{\pm}(u)=
  \frac{1}{\sin(\sfrac{\eta}{2}+\zeta\pm u)
           \sin(\sfrac{\eta}{2}-\zeta\pm u)
           +\mu\tilde{\mu}\sin^2(\eta\pm 2u)/(2\cos\eta)},
  \label{rhoK}
\ee
then $K$ satisfies
\be
  K(u)K(-u)=1,\quad K(0)=1.
  \label{KK}
\ee
\vskip.1in
\noindent {\bf Class (B):} $\sin 2\eta\neq 0$ and $\sin 4\eta=0$.
\vskip.1in
For this case we have three solutions:
\ba
  x_1(u)&\!\!\!=\!\!\!&
  \sin(\sfrac{\eta}{2}+\zeta+u)\sin(\sfrac{\eta}{2}-\zeta-u),\;
  x_2(u)=
  \sin(\sfrac{\eta}{2}+\zeta-u)\sin(\sfrac{\eta}{2}-\zeta-u),\n
  x_3(u)&\!\!\!=\!\!\!&
  \sin(\sfrac{\eta}{2}+\zeta-u)\sin(\sfrac{\eta}{2}-\zeta+u),\;
  y_1(u)=y_2(u)=\tilde{y}_1(u)=\tilde{y}_2(u)=0,
  \label{Kans2}\\
  z(u)&\!\!\!=\!\!\!&
  \nu\sin 2u,\;\tilde{z}(u)=0,\quad
  \mbox{or}\quad
  z(u)=0,\;\tilde{z}(u)=\nu\sin 2u,\nonumber
%  \label{Kans1}
\ea
with $\nu\neq 0$ and $\zeta$ ($\sin 2\zeta\neq 0$) being arbitrary
constants, and 
\ba
  y_1(u)&\!\!\!=\!\!\!&
  y_2(u)=\tilde{y}_1(u)=\tilde{y}_2(u)=x_2(u)=0,\n
  x_1(u)&\!\!\!=\!\!\!&1,\; x_3(u)=-1,\; z(u)=\nu,\; 
  \tilde{z}(u)=-\sfrac{1}{\nu},
  \label{Kans3}
\ea
where $\nu\neq 0$ is an arbitrary constant. 

If we take 
$\rho^K(u)=1/(\sin(\frac{\eta}{2}+\zeta\pm u)
\sin(\frac{\eta}{2}-\zeta\pm u))$ (four choices),
then K-matrices (\ref{Kans2}) satisfy \eq{KK}. However, K-matrix 
(\ref{Kans3}) does not satisfy \eq{KK}.

\vskip.1in
\noindent {\bf Class (C): $\sin 2\eta=0$.}
\vskip.1in
For $\eta=0,\pi$, there are no restriction for $K$.
For $\eta=\frac12\pi,\frac32\pi$,
we have $y_1=y_2=\tilde{y}_1=\tilde{y}_2=0$ and other components
are arbitrary.

%%%%%%%%%%%%%%%%%%%%%%%%%%%%%%%%%%%%%%%%%%%%%%%%%%%%%%%%%%%
%                                                         %
%  4. Spin chain Hamiltonian                              %
%                                                         %
%%%%%%%%%%%%%%%%%%%%%%%%%%%%%%%%%%%%%%%%%%%%%%%%%%%%%%%%%%%
\sect{$XXZ$ Spin-1 Chain with General Boundary}

In this section we consider the K-matrix \eq{Kanswer},
$K(u)=K(u;\zeta,\mu,\tilde{\mu})$, and take $\rho(u)=\rho_{\pm}(u)$ 
and $\rho^K(u)=\rho^K_{\pm}(u)$.
$K_{\pm}(u)$ are defined as $K_-(u)=K(u;\zeta_-,\mu_-,\tilde{\mu}_-)$,
$K_+(u)={}^tK(-u-\eta;-\zeta_+,\tilde{\mu}_+,\mu_+)$.

R- and K-matrices induce the hamiltonian of integrable 
spin-1 $XXZ$ open chain with $N$ sites,
\be
  H=
  \sum_{n=1}^{N-1}H_{n,n+1}
%  +\frac{1}{2}\frac{d}{du}\stackrel{1}{K}_-(u)\biggm|_{u=0}
  +\frac{1}{2}\frac{d}{du}K_{-,1}(u)\biggm|_{u=0}
  +\frac{\mbox{tr}_0K_{+,0}(0)H_{N,0}}{\mbox{tr}\;K_+(0)},
\ee
where the two-site hamiltonian is given by
\be
  H_{n,n+1}=\frac{d}{du}\check{R}_{n,n+1}(u)\biggm|_{u=0}.
\ee
Here suffices of $\check{R}$ and $K_{\pm}$ etc. indicate the sites on which
the operators act.
We remark that the terms in $H$ which are not proportional to identity
do not depend on the four choices ($\rho_{\pm},\rho^K_{\pm}$).
{}From \eq{R},\eq{rho},\eq{Kanswer},\eq{rhoK}, we obtain
\ba
  H&\!\!=\!\!&
  \frac{1}{\sin 2\eta}\sum_{n=1}^{N-1}\Biggl(
  \vec{s}_n\cdot\vec{s}_{n+1}-(\vec{s}_n\cdot\vec{s}_{n+1})^2
  +(1-\cos\eta)
  \Bigl\{s^3_ns^3_{n+1},s^+_ns^-_{n+1}+s^-_ns^+_{n+1}\Bigr\} \n
  &&\hspace{20mm}
  -(1-\cos 2\eta)\Bigl(s^3_ns^3_{n+1}-(s^3_n)^2(s^3_{n+1})^2
    +(s^3_n)^2+(s^3_{n+1})^2\Bigr)\Biggr) \n
  &&+\frac{1}{2\sin(\frac{\eta}{2}+\zeta_-)\sin(\frac{\eta}{2}-\zeta_-)
             +\mu_-\tilde{\mu}_-\sin\eta\tan\eta} 
  \label{Hamiltonian}\\
  &&\hspace{5mm}\times\Bigl(
  (1-\mu_-\tilde{\mu}_-)\sin\eta\:s^3_1s^3_1-\sin 2\zeta_-\:s^3_1
  +\sfrac{1}{2}\tan\eta(\mu_-^2s^+_1s^+_1+\tilde{\mu}_-^2s^-_1s^-_1) \n
  &&\hspace{10mm}
  -\sqrt{2}\sin(\sfrac{\eta}{2}+\zeta_-)
  (\mu_-s^+_1s^3_1+\tilde{\mu}_-s^3_1s^-_1)
  -\sqrt{2}\sin(\sfrac{\eta}{2}-\zeta_-)
  (\mu_-s^3_1s^+_1+\tilde{\mu}_-s^-_1s^3_1)
  \Bigr) \n
  &&+\frac{1}{2\sin(\frac{\eta}{2}+\zeta_+)\sin(\frac{\eta}{2}-\zeta_+)
             +\mu_+\tilde{\mu}_+\sin\eta\tan\eta} \n
  &&\hspace{5mm}\times\Bigl(
  (1-\mu_+\tilde{\mu}_+)\sin\eta\:s^3_Ns^3_N-\sin 2\zeta_+\:s^3_N
  +\sfrac{1}{2}\tan\eta(\mu_+^2s^+_Ns^+_N+\tilde{\mu}_+^2s^-_Ns^-_N) \n
  &&\hspace{10mm}
  -\sqrt{2}\sin(\sfrac{\eta}{2}+\zeta_+)
  (\mu_+s^+_Ns^3_N+\tilde{\mu}_+s^3_Ns^-_N)
  -\sqrt{2}\sin(\sfrac{\eta}{2}-\zeta_+)
  (\mu_+s^3_Ns^+_N+\tilde{\mu}_+s^-_Ns^3_N)
  \Bigr) \n
  && +\mbox{(constant)}\cdot\mbox{id}, \nonumber
\ea
where the spin-1 operator $s^3$, $s^{\pm}(=s^1\pm is^2)$ is given by
\be
  s^3=\left(\begin{array}{ccc}1&0&0\\0&0&0\\0&0&-1\end{array}\right),\;
  s^+=\sqrt{2}
  \left(\begin{array}{ccc}0&1&0\\0&0&1\\0&0&0\end{array}\right),\;
  s^-=
  \sqrt{2}\left(\begin{array}{ccc}0&0&0\\1&0&0\\0&1&0\end{array}\right).
\ee

In the special case, $\mu_\pm=\tilde{\mu}_\pm=0$, (\ref{Hamiltonian}) reduces
to the hamiltonian obtained in \cite{Mez90} by using the diagonal K-matrix.

%%%%%%%%%%%%%%%%%%%%%%%%%%%%%%%%%%%%%%%%%%%%%%%%%%%%%%%%%%%
%                                                         %
%  5. Fusion procedure                                    %
%                                                         %
%%%%%%%%%%%%%%%%%%%%%%%%%%%%%%%%%%%%%%%%%%%%%%%%%%%%%%%%%%%
\sect{Fusion Procedure}

The R-matrix \eq{R} of the 19-vertex model was obtained by directly 
solving the YBE  in \cite{rZF}. It can also be constructed out of the R-matrix
of the six-vertex model by using the fusion method \cite{Kul81}.

It was suggested that the fusion method can also be used to construct
K-matrices of vertex models corresponding to higher spins out of the K-matrix
of the six-vertex model \cite{Mez90}. In this section we apply this fusion
method to the present case
and compare the result derived this way with
that obtained above by directly solving the BYBE.

The R-matrix associated with the spin-$\frac12$ representation of 
$U_q(\widehat{sl_2})$ reads,
\be
R^{(\frac{1}{2},\frac{1}{2})}(u)=\rho_{\frac{1}{2}}(u)\left (
\begin{array}{cccc}
\sin(\eta+u) & 0 & 0 & 0\\
0 & \sin u & \sin\eta & 0\\
0 & \sin \eta & \sin u & 0\\
0 & 0 & 0 & \sin(\eta+u) 
\end{array}
\right ),\label{R-1/2-1/2}
\ee
with $\rho_{\frac{1}{2}}(u)\not\equiv 0$ being an arbitrary function.
The corresponding reflection K-matrix has the form \cite{deV93}
\be
K^{(\frac{1}{2})}(u)=\rho_{\frac{1}{2}}^K(u)\left (
\begin{array}{cc}
\sin(\xi+u) & \nu\sin 2u\\
\tilde{\nu}\sin 2u & \sin(\xi-u)
\end{array}
\right ),
\ee
where $\xi,\nu,\tilde{\nu}$ are constant parameters
and $\rho_\frac{1}{2}^K(u)\not\equiv 0$ is an arbitrary overall function.

According to \cite{Mez90} the K-matrix for the 19-vertex model is given
by the following fusion equation:
\be
K(u)=f(u)UP_+\left (K^{(\frac{1}{2})}(u-\sfrac{\eta}{2})\otimes 1\right )
  R^{(\frac{1}{2},\frac{1}{2})}(2u)\left (1\otimes K^{(\frac{1}{2})}(u+
  \sfrac{\eta}{2})\right )P_+U^{-1},\label{fusion-eq}
\ee
where $P_+$ is the projection onto the spin-1 component and $U$ is a
basis-changing matrix; explicitly, 
\be
P_+=\frac{1}{2}(1+P)=\left (
\begin{array}{cccc}
1 & 0 & 0 & 0\\
0 & \frac{1}{2} & \frac{1}{2} & 0\\
0 & \frac{1}{2} & \frac{1}{2} & 0\\
0 & 0 & 0 & 1
\end{array}
\right ),\quad U=\left (
\begin{array}{cccc}
1 & 0 & 0 & 0\\
0 & \alpha & \alpha & 0\\
0 & 0 & 0 & 1\\
0 & \alpha & -\alpha & 0
\end{array}
\right ).
\ee
It turns out that the K-matrix $K(u)$ from  \eq{fusion-eq} agrees with 
one of the solutions in section 3, eq.\eq{Kanswer},
if one makes the following identification:
\ba
&&f(u)=-\frac{1}{\sin(\eta+2u)}\frac{\rho^K(u)}
  {\rho_\frac{1}{2}^K(u-\frac{\eta}{2})
  \rho_\frac{1}{2}^K(u+\frac{\eta}{2})\rho_\frac{1}{2}(2u)},\n
&&\xi=\zeta,\;\nu=\frac{\mu}{\sqrt{2\cos\eta}},\;\tilde{\nu}=\frac{
\tilde{\mu}}{\sqrt{2\cos\eta}},\;\alpha=-\frac{\mu}{2\nu}.
\ea
It seems that the solutions in {\bf Class (B)} can not be obtained from this
fusion procedure.

%%%%%%%%%%%%%%%%%%%%%%%%%%%%%%%%%%%%%%%%%%%%%%%%%%%%%%%%%%%
%                                                         %
%  Acknowledgments                                        %
%                                                         %
%%%%%%%%%%%%%%%%%%%%%%%%%%%%%%%%%%%%%%%%%%%%%%%%%%%%%%%%%%%
\vskip 8mm
\noindent {\bf Acknowledgments:}
This work is supported in part by the Grant-in-Aid for Scientific
Research from the Ministry of Education, Science and Culture of
Japan, %(No.07210234, No.07740213)
and also by Japan-Korea exchange program of JSPS. Y.Z.Z. is financially
supported by JSPS.

%%%%%%%%%%%%%%%%%%%%%%%%%%%%%%%%%%%%%%%%%%%%%%%%%%%%%%%%%%%
%                                                         %
%  Appendix                                               %
%                                                         %
%%%%%%%%%%%%%%%%%%%%%%%%%%%%%%%%%%%%%%%%%%%%%%%%%%%%%%%%%%%
\appendix
\sect{Derivation of the Reflection K-matrices}

In this Appendix we sketch how \eq{Kanswer}, \eq{Kans2} and \eq{Kans3}
 are obtained and
show that there is no other nontrivial solution ({\bf Class (C)} is
trivial!). Our derivation is based on a careful case by case study.
We sometimes omit explicit expressions when
they may be lengthy.
First we remark that one can check \eq{Kanswer}, \eq{Kans2} and \eq{Kans3}
by direct substitution.
Let us denote $(i,j)$ component of \eq{BYBE} as Eq[$i,j$].
Note that Eq[$10-i,10-j$] can be obtained from Eq[$i,j$]
by interchanging $x_1\leftrightarrow x_3$,
$y_1\leftrightarrow \tilde{y}_2$,
$y_2\leftrightarrow \tilde{y}_1$,
$z\leftrightarrow \tilde{z}$.
In the following we assume $\sin 2\eta\neq 0$ (as mentioned above,
the $\sin 2\eta=0$ case is trivial).

\vskip 3mm
%%%%%%%%%%%%%
%   1       %
%%%%%%%%%%%%%
\noindent{\bf \Romannumeral{1}. }
{}From Eq[2,8], we have
\be
  \sin(u-u')\Bigl(y_1(u)y_1(u')-y_2(u)y_2(u')\Bigr)
  =
  \sin(\eta-u-u')\Bigl(y_1(u')y_2(u)-y_1(u)y_2(u')\Bigr).
  \label{a28}
\ee
%By rewriting this as
%$$
%  \frac{1-\frac{y_1}{y_2}(u)\frac{y_1}{y_2}(u')}{\sin(\eta-u-u')}
%  =
%  \frac{\frac{y_1}{y_2}(u)-\frac{y_1}{y_2}(u')}{\sin(u-u')},
%$$
Dividing this by $\sin(u-u')\sin(\eta-u-u')y_2(u)y_2(u')$ 
and taking $u'\rightarrow u$ limit, we obtain the differential equation
$$
  \frac{d}{du}\left(\frac{y_1}{y_2}\right)
  =
  \frac{1-\Bigl(\frac{y_1}{y_2}\Bigr)^2}{\sin(\eta-2u)}.
$$
Solving this, we get
\be
  \frac{y_1(u)}{y_2(u)}
  =
  \frac{\sin(\zeta-\sfrac{\eta}{2}+u)}{\sin(\zeta+\sfrac{\eta}{2}-u)},
  \label{y1/y2}
\ee
where $\zeta$ is a constant.
We can check that this relation satisfies \eq{a28}.

Eq[2,9] gives rise to
\be
  \sin(\eta-u-u')y_2(u)z(u')-\sin(\eta-2u')y_2(u')z(u)
  =
  \sin(u-u')y_1(u)z(u').
  \label{a29}
\ee
By differentiating this linear equation with respect to $u'$
and taking $u'\rightarrow u$ limit, we have
$$
  y_1=
  -\cos(\eta-2u)y_2
  +\sin(\eta-2u)y_2
  \Bigl(\frac{1}{y_2}\frac{dy_2}{du}-\frac{1}{z}\frac{dz}{du}\Bigr).
$$
Solving this differential equation with the help of \eq{y1/y2}, we obtain
\be
  \frac{y_2(u)}{z(u)}
  =
  \frac{\sin(\zeta+\sfrac{\eta}{2}-u)}{\sin(\eta-2u)}\times\mbox{(constant)}.
  \label{y2/z}
\ee
Again we can check that this relation satisfies \eq{a29}.
Up to now, we have assumed $y_2$ and $z$ are not zero.
Taking into account the fact that they may be zero,
we obtain the following result from  
\eq{a28} and \eq{a29},
\be
  y_1(u)=\mu\sin(\zeta-\sfrac{\eta}{2}+u)g(u),\;
  y_2(u)=\mu\sin(\zeta+\sfrac{\eta}{2}-u)g(u),\;
  z(u)=\nu\sin(\eta-2u)g(u),
  \label{y1y2z}
\ee
where $\zeta$, $\mu$, $\nu$ are arbitrary constants and 
$g(u)\not\equiv 0$ is an arbitrary function.

Similarly, by  Eq[8,2] and Eq[9,2],
\be
  \tilde{y}_1(u)=\tilde{\mu}'\sin(\tilde{\zeta}-\sfrac{\eta}{2}+u)
  \tilde{g}(u),\;
  \tilde{y}_2(u)=\tilde{\mu}'\sin(\tilde{\zeta}+\sfrac{\eta}{2}-u)
  \tilde{g}(u),\;
  \tilde{z}(u)=\tilde{\nu}'\sin(\eta-2u)\tilde{g}(u),%\n
  \label{ty1ty2tz}
\ee
where $\tilde{\zeta}$, $\tilde{\mu}'$, $\tilde{\nu}'$ are arbitrary
constants and $\tilde{g}(u)\not\equiv 0$ is an arbitrary function.

{}From Eq[1,1] and Eq[9,9], we have
\ba
  \sin(\eta+u+u')\Bigl(\tilde{y}_1(u')y_1(u)-\tilde{y}_1(u)y_1(u')\Bigr)
  &\!\!=\!\!&
  \sin\eta\Bigl(z(u')\tilde{z}(u)-z(u)\tilde{z}(u')\Bigr),
  \label{a11}\\
  \sin(\eta+u+u')\Bigl(\tilde{y}_2(u')y_2(u)-\tilde{y}_2(u)y_2(u')\Bigr)
  &\!\!=\!\!&
  \sin\eta\Bigl(z(u')\tilde{z}(u)-z(u)\tilde{z}(u')\Bigr).
  \label{a99}
\ea
Using \eq{y1y2z} and \eq{ty1ty2tz},
we  arrive at the following three partial results:
\ba
  &(1)&
  \left\{
  \begin{array}{rclrcl}
    y_1(u)&\!\!=\!\!&\mu\sin(\zeta-\sfrac{\eta}{2}+u)g(u),&
    \tilde{y}_1(u)&\!\!=\!\!&\tilde{\mu}\sin(\zeta-\sfrac{\eta}{2}+u)g(u),\\
    y_2(u)&\!\!=\!\!&\mu\sin(\zeta+\sfrac{\eta}{2}-u)g(u),&
    \tilde{y}_2(u)&\!\!=\!\!&\tilde{\mu}\sin(\zeta+\sfrac{\eta}{2}-u)g(u),\\
    z(u)&\!\!=\!\!&\nu\sin(\eta-2u)g(u),&
    \tilde{z}(u)&\!\!=\!\!&\tilde{\nu}\sin(\eta-2u)g(u),
  \end{array}
  \right.\\
  &(2)&
  \left\{
  \begin{array}{rclrcl}
    y_1(u)&\!\!=\!\!&\sin(\zeta-\sfrac{\eta}{2}+u)g(u),&
    \tilde{y}_1(u)&\!\!=\!\!&0,\\
    y_2(u)&\!\!=\!\!&\sin(\zeta+\sfrac{\eta}{2}-u)g(u),&
    \tilde{y}_2(u)&\!\!=\!\!&0,\\
    z(u)&\!\!=\!\!&0,&
    \tilde{z}(u)&\!\!=\!\!&\sin(\eta-2u)\tilde{g}(u),
  \end{array}
  \right.\\
  &(3)&
  \left\{
  \begin{array}{rclrcl}
    y_1(u)&\!\!=\!\!&0,&
    \tilde{y}_1(u)&\!\!=\!\!&\sin(\tilde{\zeta}-\sfrac{\eta}{2}+u)
    \tilde{g}(u),\\
    y_2(u)&\!\!=\!\!&0,&
    \tilde{y}_2(u)&\!\!=\!\!&\sin(\tilde{\zeta}+\sfrac{\eta}{2}-u)
    \tilde{g}(u),\\
    z(u)&\!\!=\!\!&\sin(\eta-2u)g(u),&
    \tilde{z}(u)&\!\!=\!\!&0,
  \end{array}
  \right.
\ea
where $\zeta$, $\tilde{\zeta}$, $\mu$, $\tilde{\mu}$, $\nu$, 
$\tilde{\nu}$ are arbitrary constants and $g(u)$ and $\tilde{g}(u)$ are
arbitrary functions ($g,\tilde{g}\not\equiv 0$, $g/\tilde{g}\not\equiv
\mbox{constant}$).

\vskip 3mm
%%%%%%%%%%%%%
%   2       %
%%%%%%%%%%%%%
\noindent{\bf \Romannumeral{2}. }
First let us consider case (2).
Eq[4,1] and Eq[2,1] can be written as the following form respectively
\ban
  A_1(u,u')g(u)\tilde{g}(u')+A_2(u,u')g(u')\tilde{g}(u)
  &\!\!=\!\!& 0, \\
  A_3(u,u')g(u)\tilde{g}(u')+A_4(u,u')g(u')\tilde{g}(u)
  &\!\!=\!\!& 0,
\ean
where $A_i$'s are known functions of $u$ and $u'$ with their 
explicit expressions omitted.
Since $g(u)\tilde{g}(u')\not\equiv 0$, we have $\Delta(u,u')\equiv
A_1A_4-A_2A_3=0$. $\Delta(0,\eta)=0$ means $\sin 2\zeta=0$.
With this $\zeta$, however, $\Delta(u,-u)$ does not vanish, which implies
the above two equations are contradictory each other.
Therefore case (2) is not a solution.

Similarly, from Eq[1,4] and Eq[1,2],  we can show that 
case (3)  is not a solution.

\vskip 3mm
%%%%%%%%%%%%%
%   3       %
%%%%%%%%%%%%%
\noindent{\bf \Romannumeral{3}. }
In the following we consider case (1).
Eq[1,4] is linear in $x_1(u),x_1(u'),x_2(u),x_2(u')$.
There are two cases to be considered: $\mu=0$ and $\mu\neq 0$.
For the former case, Eq[1,4] is satisfied if and only if 
$\tilde{\mu}\nu=0$. For the latter case, the coefficient of $x_2(u')$ 
vanishes and $x_2(u)$ can be expressed by $x_1(u)$ and $x_1(u')$.
Of course this expression must be independent on $u'$.
Similar results can be obtained for Eq[4,1].
Combining these results, we derive
\ba
  &{\rm (\romannumeral1)}&
   \mu=0,\tilde{\mu}=0  \n
  &{\rm (\romannumeral2)}&
   \mu=0,\tilde{\mu}\neq 0 \qquad(\Rightarrow \nu=0) \n
  &{\rm (\romannumeral3)}&
   \mu\neq 0,\tilde{\mu}=0 \qquad(\Rightarrow \tilde{\nu}=0) \\
  &{\rm (\romannumeral4\!\!-\!\!a)}&
   \mu\neq 0,\tilde{\mu}\neq 0, \nu=0 \qquad
  (\Rightarrow \tilde{\nu}=0) \n
  &{\rm (\romannumeral4\!\!-\!\!b)}&
   \mu\neq 0,\tilde{\mu}\neq 0, \nu\neq 0 \qquad
  \biggl(\Rightarrow 
  \tilde{\nu}=\Bigl(\frac{\tilde{\mu}}{\mu}\Bigr)^2\nu,\,
  \sin(\eta\mp 2\zeta)\neq 0\biggr) \nonumber
\ea
(the condition $\sin(\eta+2\zeta)\neq 0$ is obtained below).
For (\romannumeral1) there is no restriction on $x_1,x_2$.
For other cases,
\ba
  x_1(u)&\!\!=\!\!&
  \biggl(\frac{\lambda_1}{\sin u}+\frac{\lambda_2}{\cos u}\biggr)
  \sin(\sfrac{\eta}{2}-\zeta-u)g(u)
  +\delta\frac{\tilde{\mu}\nu}{\mu}
  \frac{\sin\eta\sin2\zeta}{\sin(\eta-2\zeta)}g(u),
  \label{x1x2} \\
  x_2(u)&\!\!=\!\!&
  \biggl(\frac{\lambda_1}{\sin u}-\frac{\lambda_2}{\cos u}\biggr)
  \sin(\sfrac{\eta}{2}-\zeta-u)g(u)
  +\delta\frac{\tilde{\mu}\nu}{\mu}
  \biggl(\frac{\sin^2\eta}{\sin(\eta-2\zeta)}-\sin 2u\biggr)g(u),
  \nonumber
\ea
where $\lambda_1,\lambda_2$ are arbitrary constants, and 
$\delta=1$ for case (\romannumeral4-b) and
$\delta=0$ for other cases.

Similarly from Eq[9,6] and Eq[6,9] we obtain
$x_3$ and $x_2$ (and the condition $\sin(\eta+2\zeta)\neq 0$ in the
case (\romannumeral4-b) above): for
(\romannumeral1) there is no restriction on $x_3,x_2$, and for other cases,
\ba
  x_3(u)&\!\!=\!\!&
  \biggl(\frac{\lambda_3}{\sin u}+\frac{\lambda_4}{\cos u}\biggr)
  \sin(\sfrac{\eta}{2}+\zeta-u)g(u)
  -\delta\frac{\tilde{\mu}\nu}{\mu}
  \frac{\sin\eta\sin2\zeta}{\sin(\eta+2\zeta)}g(u),
  \label{x3x2} \\
  x_2(u)&\!\!=\!\!&
  \biggl(\frac{\lambda_3}{\sin u}-\frac{\lambda_4}{\cos u}\biggr)
  \sin(\sfrac{\eta}{2}+\zeta-u)g(u)
  +\delta\frac{\tilde{\mu}\nu}{\mu}
  \biggl(\frac{\sin^2\eta}{\sin(\eta+2\zeta)}-\sin 2u\biggr)g(u),
  \nonumber
\ea
where $\lambda_3,\lambda_4$ are arbitrary constants, and 
$\delta=1$ for case (\romannumeral4-b) and
$\delta=0$ for other cases.

Combining \eq{x1x2} and \eq{x3x2} (for example taking the limit 
$u\rightarrow 0,\frac{\pi}{2}$), we get
\ba
  x_1(u)&\!\!=\!\!&
  \frac{2\lambda}{\sin 2u}\sin(\sfrac{\eta}{2}+\zeta+u)
  \sin(\sfrac{\eta}{2}-\zeta-u)g(u) \n
  &&
  +\delta\frac{\tilde{\mu}\nu}{\mu}\frac{\sin\eta}{\sin(\eta-2\zeta)}
  \biggl(\sin 2\zeta-\frac{\sin(\sfrac{\eta}{2}-\zeta-u)\sin 2\eta}
  {2\cos u\sin(\sfrac{\eta}{2}+\zeta)}\biggr)g(u), \n
  x_2(u)&\!\!=\!\!&
  \frac{2\lambda}{\sin 2u}\sin(\sfrac{\eta}{2}+\zeta-u)
  \sin(\sfrac{\eta}{2}-\zeta-u)g(u) \\
  &&
  +\delta\frac{\tilde{\mu}\nu}{\mu}\frac{\sin\eta}{\sin(\eta-2\zeta)}
  \biggl(\sin\zeta+\frac{\sin(\sfrac{\eta}{2}-\zeta-u)\sin 2\eta}
  {2\cos u\sin(\sfrac{\eta}{2}+\zeta)}\biggr)g(u)
  -\delta\frac{\tilde{\mu}\nu}{\mu}\sin 2u\,g(u), \n
  x_3(u)&\!\!=\!\!&
  \frac{2\lambda}{\sin 2u}\sin(\sfrac{\eta}{2}-\zeta+u)
  \sin(\sfrac{\eta}{2}+\zeta-u)g(u) \n
  &&
  +\delta\frac{\tilde{\mu}\nu}{\mu}\frac{\sin\eta}{\sin(\eta+2\zeta)}
  \biggl(-\sin 2\zeta-\frac{\sin(\sfrac{\eta}{2}+\zeta-u)\sin 2\eta}
  {2\cos u\sin(\sfrac{\eta}{2}-\zeta)}\biggr)g(u), \nonumber
\ea
where $\delta=1$ for case (\romannumeral4-b) and
$\delta=0$ for cases 
(\romannumeral2),(\romannumeral3),(\romannumeral4-a).
For case (\romannumeral1) there are no restriction for $x_1,x_2,x_3$.

\vskip 3mm
%%%%%%%%%%%%%
%   4       %
%%%%%%%%%%%%%
\noindent{\bf \Romannumeral{4}. }
Case (\romannumeral1).\\
Let $X_i(u)=x_i(u)/(\sin(\eta-2u)g(u))$. 
Eq[4,6] and Eq[6,4]  lead to
\ba
  && \nu\Bigl(X_1(u)-X_3(u)-(X_1(u')-X_3(u'))\Bigr)=0,
  \label{a46} \\
  && \tilde{\nu}\Bigl(X_1(u)-X_3(u)-(X_1(u')-X_3(u'))\Bigr)=0,
  \label{a64}
\ea
and Eq[2,4] and Eq[6,8] to
\ba
  \frac{X_1(u)X_1(u')-X_2(u)X_2(u')+\nu\tilde{\nu}}{\sin(u+u')}
  &\!\!=\!\!&
  \frac{X_1(u)X_2(u')-X_1(u')X_2(u)}{\sin(u-u')},
  \label{a24} \\
  \frac{X_3(u)X_3(u')-X_2(u)X_2(u')+\nu\tilde{\nu}}{\sin(u+u')}
  &\!\!=\!\!&
  \frac{X_3(u)X_2(u')-X_3(u')X_2(u)}{\sin(u-u')}.
  \label{a68}
\ea
We have two choices (a) $\nu\tilde{\nu}=0$, (b) $\nu\tilde{\nu}\neq 0$.

%%%%%%%%%%%%%
%   4.1     %
%%%%%%%%%%%%%
\noindent{\bf \Romannumeral{4}.1. }
Case (a).\\
Through a similar calculation as for \eq{a28}, it follows 
from \eq{a24} and \eq{a68} that for $x_2\equiv 0$, we have
$x_1(u)=x_2(u)=x_3(u)=0$, and for $x_2\not\equiv 0$, we get 
\ignore{
\be
  \frac{x_1(u)}{x_2(u)}=\frac{\sin(c_1-u)}{\sin(c_1+u)},\quad
  \frac{x_3(u)}{x_2(u)}=\frac{\sin(c_2-u)}{\sin(c_2+u)},
\ee
}
\ba
  x_1(u)&\!\!=\!\!&\sin(c_1-u)\sin(c_2+u)f(u),\n
  x_2(u)&\!\!=\!\!&\sin(c_1+u)\sin(c_2+u)f(u),\\
  x_3(u)&\!\!=\!\!&\sin(c_1+u)\sin(c_2-u)f(u),\nonumber
\ea
where $c_1,c_2$ are arbitrary constants and 
$f(u)\not\equiv 0$ is an arbitrary function.
For the former case, we can check that they give the following two
solutions of \eq{BYBE}:
%\ba
%&{\rm (s1)}&  z(u)=h(u),\:
%  x_1(u)=x_2(u)=x_3(u)=0, \n
%& &y_1(u)=y_2(u)=\tilde{y}_1(u)=\tilde{y}_2(u)=
%  \tilde{z}(u)=0;
%  \label{sol1}\\
%&{\rm (s2)}&  \tilde{z}(u)=h(u),\:
%  x_1(u)=x_2(u)=x_3(u)=0, \n
%& &y_1(u)=y_2(u)=\tilde{y}_1(u)=\tilde{y}_2(u)=
%  z(u)=0.
%  \label{sol2}
%\ea
\ba
  && z(u)=h(u),\:
  x_1=x_2=x_3=y_1=y_2=\tilde{y}_1=\tilde{y}_2=\tilde{z}=0,
  \label{sol1}\\
  && \tilde{z}(u)=h(u),\:
  x_1=x_2=x_3=y_1=y_2=\tilde{y}_1=\tilde{y}_2=z=0,
  \label{sol2}
\ea
where $h(u)\not\equiv 0$ is an arbitrary function.
For the latter case, Eq[3,5] implies $\sin(c_1+c_2+\eta)=0$.
Since $x_i$'s are invariant under $c_2\rightarrow c_2+\pi$ and 
$f(u)\rightarrow -f(u)$, we can take $c_1+c_2+\eta=0$.
Therefore we have
\ba
  x_1(u)&\!\!=\!\!&
  \sin(\sfrac{\eta}{2}+\zeta+u)\sin(\sfrac{\eta}{2}-\zeta-u)f(u),\n
  x_2(u)&\!\!=\!\!&
  \sin(\sfrac{\eta}{2}+\zeta-u)\sin(\sfrac{\eta}{2}-\zeta-u)f(u),
  \label{x1x2x3}\\
  x_3(u)&\!\!=\!\!&
  \sin(\sfrac{\eta}{2}+\zeta-u)\sin(\sfrac{\eta}{2}-\zeta+u)f(u),
  \nonumber
\ea
where $\zeta$ is an arbitrary constant. The following three cases are possible:
\ba
  \mbox{(a1)}&&\nu=\tilde{\nu}=0,\n
  \mbox{(a2)}&&\nu\neq 0,\:\tilde{\nu}=0,\\
  \mbox{(a3)}&&\nu=0,\:\tilde{\nu}\neq 0. \nonumber
\ea
\noindent
$\bullet$ Case (a1).
We can check that \eq{BYBE} is satisfied.
This is a diagonal solution,
\ba
  y_1(u)&\!\!\!=\!\!\!&
  y_2(u)=\tilde{y}_1(u)=\tilde{y}_2(u)=z(u)=\tilde{z}(u)=0, \n
  x_1(u)&\!\!\!=\!\!\!&
  \sin(\sfrac{\eta}{2}+\zeta+u)\sin(\sfrac{\eta}{2}-\zeta-u)f(u),\n
  x_2(u)&\!\!\!=\!\!\!&
  \sin(\sfrac{\eta}{2}+\zeta-u)\sin(\sfrac{\eta}{2}-\zeta-u)f(u),
  \label{sol3}\\
  x_3(u)&\!\!\!=\!\!\!&
  \sin(\sfrac{\eta}{2}+\zeta-u)\sin(\sfrac{\eta}{2}-\zeta+u)f(u),
  \nonumber
\ea
where $\zeta$ is an arbitrary constant and $f(u)\not\equiv 0$ is an 
arbitrary function.

\noindent
$\bullet$ Case (a2).
{}From \eq{a46} we have $x_1(u)-x_3(u)=\sigma\sin(\eta-2u)g(u)$,
where $\sigma$ is a constant.
{}From \eq{x1x2x3} we get $-\sin 2\zeta\sin 2uf(u)=
\sigma\sin(\eta-2u)g(u)$. There are two choices:
\ba
  \mbox{(a2-1)}&&\sin 2\zeta\neq 0 \: (\Rightarrow\sigma\neq 0),\n
  \mbox{(a2-2)}&&\sin 2\zeta=0\: (\Rightarrow\sigma=0).
\ea
For the case (a2-1), Eq[1,3] implies $\sin 4\eta=0$ and then we can check
that \eq{BYBE} is satisfied. Therefore we get a solution (note: 
$\sin 4\eta=0$),
\ba
  z(u)&\!\!\!=\!\!\!&\nu\sin 2u\:h(u),\quad
  y_1(u)=y_2(u)=\tilde{y}_1(u)=\tilde{y}_2(u)=\tilde{z}(u)=0, \n
  x_1(u)&\!\!\!=\!\!\!&
  \sin(\sfrac{\eta}{2}+\zeta+u)\sin(\sfrac{\eta}{2}-\zeta-u)h(u),\n
  x_2(u)&\!\!\!=\!\!\!&
  \sin(\sfrac{\eta}{2}+\zeta-u)\sin(\sfrac{\eta}{2}-\zeta-u)h(u),
  \label{sol4}\\
  x_3(u)&\!\!\!=\!\!\!&
  \sin(\sfrac{\eta}{2}+\zeta-u)\sin(\sfrac{\eta}{2}-\zeta+u)h(u),
  \nonumber
\ea
where $\nu\neq 0$ and $\zeta$ ($\sin 2\zeta\neq 0$) are arbitrary
constants and $h(u)\not\equiv 0$ is an arbitrary function.

For the case (a2-2), we may take $\zeta=0$ or $\frac12\pi$ 
because \eq{x1x2x3} is
periodic in $\zeta$ with period $\pi$. For both the $\zeta$ values,  Eq[2,6]
implies $\sin(\eta-2u)g(u)=(\mbox{constant})\times\sin 2uf(u)$.
However Eq[1,3] implies this constant is zero, which contradicts 
$g(u)\not\equiv 0$.

\noindent
$\bullet$ Case (a3).
{}From \eq{a64} we have $x_1(u)-x_3(u)=\sigma\sin(\eta-2u)g(u)$,
where $\sigma$ is a constant.
{}From \eq{x1x2x3} we get $-\sin 2\zeta\sin 2uf(u)=
\sigma\sin(\eta-2u)g(u)$. Again two choices are possible:
\ba
  \mbox{(a3-1)}&&\sin 2\zeta\neq 0\: (\Rightarrow\sigma\neq 0),\n
  \mbox{(a3-2)}&&\sin 2\zeta=0\: (\Rightarrow\sigma=0).
\ea
For the case (a2-1), Eq[3,1] implies $\sin 4\eta=0$ and then we can check
that \eq{BYBE} is satisfied. Therefore we get a solution (note again: 
$\sin 4\eta=0$),
\ba
  \tilde{z}(u)&\!\!\!=\!\!\!&\tilde{\nu}\sin 2u\:h(u),\quad
  y_1(u)=y_2(u)=\tilde{y}_1(u)=\tilde{y}_2(u)=z(u)=0, \n
  x_1(u)&\!\!\!=\!\!\!&
  \sin(\sfrac{\eta}{2}+\zeta+u)\sin(\sfrac{\eta}{2}-\zeta-u)h(u),\n
  x_2(u)&\!\!\!=\!\!\!&
  \sin(\sfrac{\eta}{2}+\zeta-u)\sin(\sfrac{\eta}{2}-\zeta-u)h(u),
  \label{sol5}\\
  x_3(u)&\!\!\!=\!\!\!&
  \sin(\sfrac{\eta}{2}+\zeta-u)\sin(\sfrac{\eta}{2}-\zeta+u)h(u),
  \nonumber
\ea
where $\tilde{\nu}\neq 0$ and $\zeta$ ($\sin 2\zeta\neq 0$) are
arbitrary constants and $h(u)\not\equiv 0$ is an arbitrary function.
For the case (a3-2), 
%we take $\zeta=0,\pi/2$ because \eq{x1x2x3} is
%periodic in $\zeta$ with period $\pi$. For both cases 
Eq[6,2]
implies $\sin(\eta-2u)g(u)=(\mbox{constant})\times\sin 2uf(u)$.
However Eq[3,1] implies this constant is zero, which contradicts 
$g(u)\not\equiv 0$.

%%%%%%%%%%%%%
%   4.2     %
%%%%%%%%%%%%%
\noindent{\bf \Romannumeral{4}.2. }
Case (b).\\
\eq{a24} with $u'=0$ implies that $X_1(u)+X_2(u)$ is a constant,
and \eq{a24} with $u'=\frac12\pi$ implies that $X_1(u)-X_2(u)$ is also 
a constant. Therefore $X_1$ and $X_2$ are both constants, 
$X_1(u)=c_1,X_2(u)=c_2$ with $c_1^2-c_2^2+\nu\tilde{\nu}=0$.
Similarly from \eq{a68} we have $X_3(u)=c_3$ with 
$c_3^2-c_2^2+\nu\tilde{\nu}=0$.
We have two possibilities, $c_3=c_1$ and $c_3=-c_1$. 
We eliminate $\tilde{\nu}$ using $\tilde{\nu}=(c_2^2-c_1^2)/\nu$.

\noindent
$\bullet$ Case $c_3=c_1$.
Eq[2,6] implies $c_1=0$ and Eq[1,3] implies 
$c_2=0$. But this contradicts $\tilde{\nu}\neq 0$.

\noindent
$\bullet$ Case $c_3=-c_1$.
Eq[3,7] implies $c_1=0$ or $c_2=0$.
For $c_1=0$, Eq[1,3] implies $c_2=0$ but this contradicts 
$\tilde{\nu}\neq 0$.
For $c_2=0$, Eq[1,3] implies $c_1=0$ or $\sin 4\eta=0$. 
The former contradicts $\tilde{\nu}\neq 0$. For the latter case
we can check \eq{BYBE} is satisfied.
Therefore we get a solution (for $\sin 4\eta=0$),
\ba
  && y_1(u)=y_2(u)=\tilde{y}_1(u)=\tilde{y}_2(u)=x_2(u)=0,\n
  && x_1(u)=h(u),\: x_3(u)=-h(u),\: z(u)=\nu h(u),\: 
     \tilde{z}(u)=-\sfrac{1}{\nu}h(u),
  \label{sol6}
\ea
where $\nu\neq 0$ is an arbitrary constant and $h(u)\not\equiv 0$ is
an arbitrary function.

\vskip 3mm
%%%%%%%%%%%%%
%   5       %
%%%%%%%%%%%%%
\noindent{\bf \Romannumeral{5}. }
Case (\romannumeral2), (\romannumeral3)\\
First let us consider case (\romannumeral2).
Eq[6,2] implies $\tilde{\mu}^2=4\lambda\tilde{\nu}\cos\eta$.
Then we can check that \eq{BYBE} is satisfied.
Therefore we get a solution,
\ba
  y_1(u)&\!\!\!=\!\!\!& y_2(u)=z(u)=0,\n
  \tilde{y}_1(u)&\!\!\!=\!\!\!&
  \tilde{\mu}\sin(\zeta-\sfrac{\eta}{2}+u)g(u),\;\;
  \tilde{y}_2(u)=\tilde{\mu}\sin(\zeta+\sfrac{\eta}{2}-u)g(u),\;\;
  \tilde{z}(u)=\tilde{\nu}\sin(\eta-2u)g(u),\n
  x_1(u)&\!\!\!=\!\!\!&\frac{\tilde{\mu}^2}{2\tilde{\nu}\cos\eta\sin 2u}
  \sin(\sfrac{\eta}{2}+\zeta+u)\sin(\sfrac{\eta}{2}-\zeta-u)g(u),
  \label{sol7} \\
  x_2(u)&\!\!\!=\!\!\!&\frac{\tilde{\mu}^2}{2\tilde{\nu}\cos\eta\sin 2u}
  \sin(\sfrac{\eta}{2}+\zeta-u)\sin(\sfrac{\eta}{2}-\zeta-u)g(u), \n
  x_3(u)&\!\!\!=\!\!\!&\frac{\tilde{\mu}^2}{2\tilde{\nu}\cos\eta\sin 2u}
  \sin(\sfrac{\eta}{2}+\zeta-u)\sin(\sfrac{\eta}{2}-\zeta+u)g(u),
  \nonumber
\ea
where $\zeta,\tilde{\mu},\tilde{\nu}$ are arbitrary constant
($\tilde{\mu}\neq 0$, $\tilde{\nu}\neq 0$) and $g(u)\not\equiv 0$ is
an arbitrary function.

Case (\romannumeral3) is treated similarly: Eq[2,6] implies 
$\mu^2=4\lambda\nu\cos\eta$ and we get a solution,
\ba
  y_1(u)&\!\!\!=\!\!\!&\mu\sin(\zeta-\sfrac{\eta}{2}+u)g(u),\;\;
  y_2(u)=\mu\sin(\zeta+\sfrac{\eta}{2}-u)g(u),\;\;
  z(u)=\nu\sin(\eta-2u)g(u),\n
  \tilde{y}_1(u)&\!\!\!=\!\!\!& \tilde{y}_2(u)=\tilde{z}(u)=0,\n
  x_1(u)&\!\!\!=\!\!\!&\frac{\mu^2}{2\nu\cos\eta\sin 2u}
  \sin(\sfrac{\eta}{2}+\zeta+u)\sin(\sfrac{\eta}{2}-\zeta-u)g(u),
  \label{sol8}\\
  x_2(u)&\!\!\!=\!\!\!&\frac{\mu^2}{2\nu\cos\eta\sin 2u}
  \sin(\sfrac{\eta}{2}+\zeta-u)\sin(\sfrac{\eta}{2}-\zeta-u)g(u), \n
  x_3(u)&\!\!\!=\!\!\!&\frac{\mu^2}{2\nu\cos\eta\sin 2u}
  \sin(\sfrac{\eta}{2}+\zeta-u)\sin(\sfrac{\eta}{2}-\zeta+u)g(u),
  \nonumber
\ea
where $\zeta,\mu,\nu$ are arbitrary constant
($\mu\neq 0$, $\nu\neq 0$) and $g(u)\not\equiv 0$ is an arbitrary
function.

\vskip 3mm
%%%%%%%%%%%%%
%   6       %
%%%%%%%%%%%%%
\noindent{\bf \Romannumeral{6}. }
Case (\romannumeral4-a)\\
Eq[1,7] implies $\mu=0$ and Eq[7,1] implies $\tilde{\mu}=0$.
But this contradicts the assumption. Therefore there are no solutions
for this case.

\vskip 3mm
%%%%%%%%%%%%%
%   7       %
%%%%%%%%%%%%%
\noindent{\bf \Romannumeral{7}. }
Case (\romannumeral4-b)\\
Eq[2,4] implies 
\be
  \lambda=
  \frac{\mu^3\sin(\sfrac{\eta}{2}+\zeta)\sin(\sfrac{\eta}{2}-\zeta)
        +\tilde{\mu}\nu^2\sin\eta\sin 2\eta}
       {4\mu\nu\cos\eta
        \sin(\sfrac{\eta}{2}+\zeta)\sin(\sfrac{\eta}{2}-\zeta)}.
\ee
Then we can check that \eq{BYBE} is satisfied.
Therefore we get a solution,
\ba
  y_1(u)&\!\!\!=\!\!\!&\mu\sin(\zeta-\sfrac{\eta}{2}+u)g(u),\;\;
  y_2(u)=\mu\sin(\zeta+\sfrac{\eta}{2}-u)g(u),\;\;
  z(u)=\nu\sin(\eta-2u)g(u),\n
  \tilde{y}_1(u)&\!\!\!=\!\!\!&
  \tilde{\mu}\sin(\zeta-\sfrac{\eta}{2}+u)g(u),\;\;
  \tilde{y}_2(u)=\tilde{\mu}\sin(\zeta+\sfrac{\eta}{2}-u)g(u),\;\;
  \tilde{z}(u)=\tilde{\nu}\sin(\eta-2u)g(u),\n
  x_1(u)&\!\!\!=\!\!\!&\frac{
  \mu^3\sin(\sfrac{\eta}{2}+\zeta+u)\sin(\sfrac{\eta}{2}-\zeta-u)
  +\tilde{\mu}\nu^2\sin 2\eta\sin(\eta-2u)}{2\mu\nu\cos\eta\sin 2u}g(u),
  \label{sol9}\\
  x_2(u)&\!\!\!=\!\!\!&\frac{
  \mu^3\sin(\sfrac{\eta}{2}+\zeta-u)\sin(\sfrac{\eta}{2}-\zeta-u)
  +2\tilde{\mu}\nu^2\cos\eta\sin(\eta-2u)\sin(\eta+2u)}
  {2\mu\nu\cos\eta\sin 2u}
  g(u), \n
  x_3(u)&\!\!\!=\!\!\!&\frac{
  \mu^3\sin(\sfrac{\eta}{2}+\zeta-u)\sin(\sfrac{\eta}{2}-\zeta+u)
  +\tilde{\mu}\nu^2\sin 2\eta\sin(\eta-2u)}
  {2\mu\nu\cos\eta\sin 2u}g(u), \nonumber
\ea
where $\zeta,\mu,\tilde{\mu},\nu$ are arbitrary constants 
($\mu\neq 0$, $\tilde{\mu}\neq 0$, $\nu\neq 0$, 
$\tilde{\nu}=\tilde{\mu}^2\nu/\mu^2$)
and $g(u)\not\equiv 0$ is an arbitrary function.

\vskip 3mm
%%%%%%%%%%%%%
%   8       %
%%%%%%%%%%%%%
\noindent{\bf \Romannumeral{8}. }
To summarize, we have obtained all the solutions of BYBE.
%\eq{sol1},\eq{sol2},\eq{sol3},\eq{sol4},\eq{sol5},
%\eq{sol6},\eq{sol7},\eq{sol8},\eq{sol9}.
\eq{sol4}, \eq{sol5}, \eq{sol6} correspond to 
three solutions in {\bf Class (B)}.
Other solutions 
\eq{sol1}, \eq{sol2}, \eq{sol3}, \eq{sol7}, \eq{sol8}, \eq{sol9}
can be expressed in one formula \eq{Kanswer} ({\bf Class (A)}).
For example, \eq{sol9} is obtained from \eq{Kanswer} by 
replacing $\mu\rightarrow 2\nu\mu^{-1}\cos\eta$, 
$\tilde{\mu}\rightarrow 2\tilde{\mu}\nu\mu^{-2}\cos\eta$
and dividing by $2\nu\mu^{-2}\cos\eta\sin 2u$.
\eq{sol1} is obtained from \eq{Kanswer} by 
setting $\tilde{\mu}=0$, dividing by $\mu^2$, and taking a
limit $\mu\rightarrow\infty$.
But this solution can not satisfy \eq{KK}.

%%%%%%%%%%%%%%%%%%%%%%%%%%%%%%%%%%%%%%%%
%                                      %
%   References                         %
%                                      %
%%%%%%%%%%%%%%%%%%%%%%%%%%%%%%%%%%%%%%%%

\end{document}